\documentclass[letter]{aa}  
\usepackage{graphicx}
\usepackage{multirow}
\usepackage{txfonts}
\usepackage[]{hyperref}

\begin{document} 
   \title{Zooming into the neutrino-associated blazar candidate J1718+4239}
   \author{K. Perger\inst{1,2}         
           \and
          S. Frey\inst{1,2,3}
          \and
          K. \'{E}. Gab\'{a}nyi\inst{4,5,1,2,6}
          \and
          E. Kun\inst{7,8,9,1,2}
          }
   \institute{
             Konkoly Observatory, HUN-REN Research Centre for Astronomy and Earth Sciences, Konkoly-Thege Mikl\'os \'ut 15-17, 1121 Budapest, Hungary  
              \email{perger.krisztina@csfk.org}             
    \and    CSFK, MTA Centre of Excellence, Konkoly-Thege Mikl\'os \'ut 15-17, 1121 Budapest, Hungary   
    \and    Institute of Physics and Astronomy, ELTE E\"otv\"os Lor\'and University, P\'azm\'any P\'eter s\'et\'any 1/A, 1117 Budapest, Hungary
    \and Department of  Astronomy, Institute of Physics and Astronomy, ELTE E\"otv\"os Lor\'and University, P\'azm\'any P\'eter s\'et\'any 1/A, 1117 Budapest, Hungary
    \and   HUN-REN--ELTE Extragalactic Astrophysics Research Group, ELTE E\"otv\"os Lor\'and University, P\'azm\'any P\'eter s\'et\'any 1/A, 1117 Budapest, Hungary   
    \and
    Institute of Astronomy, Faculty of Physics, Astronomy and Informatics, Nicolaus Copernicus University, Grudzi\c{a}dzka 5, 87-100 Toru\'n, Poland
    \and Theoretical Physics IV, Faculty for Physics \& Astronomy, Ruhr University Bochum, D-44780 Bochum, Germany
    \and Faculty for Physics \& Astronomy, Astronomical Institute, Ruhr University Bochum, D-44780 Bochum, Germany
    \and Ruhr Astroparticle and Plasma Physics Center, Ruhr-Universit\"at Bochum, D-44780 Bochum, Germany 
    }

   \date{Received April 16, 2025; accepted May 9, 2025}
 
  \abstract{The radio quasar NVSS~J171822+423948 (J1718+4239) was proposed as the counterpart of the IceCube neutrino event IC-201221A. To reveal the nature of the source,  we conducted new very long baseline interferometry (VLBI) observations of this blazar candidate with the Very Long Baseline Array (VLBA). The observations were carried out in dual-band mode between $4$ and $7$~GHz. Archival radio data from the literature were also collected for comparison. Our analysis revealed highly variable, Doppler-boosted radio emission of the source, with compact structure at both kpc and pc scales, a slightly inverted spectrum, and a maximum jet inclination angle of $\theta\le5\degr$. These results confirm J1718+4239 as a blazar-type object.}

   \maketitle

\section{Introduction}\label{sec:intro}

To date, the exact astrophysical process behind the origin of cosmic high-energy neutrinos is still uncertain. Different astrophysical objects and scenarios are discussed in the literature, e.g. X-ray binary systems \citep[e.g.][]{
2005NewAR..49....1B,2008MNRAS.387.1745R,2024A&A...690A..87K}, superbubbles surrounding regions of massive star formation \citep[e.g.][]{2022ApJ...931L..30B}, novae \citep[e.g.][]{2022MNRAS.511.3339B,
2023ApJ...953..160A}, supernovae \citep[e.g.][]{2005NewAR..49....1B,2015JCAP...06..034B,2018ApJ...856..119H}, starburst galaxies \citep[e.g.][]{
2020PhRvD.102b3034A,2021PhRvD.104h3013A}, the Galactic plane \citep{2023Sci...380.1338I}, and active galactic nuclei (AGN) through various processes \citep[e.g.][]{2025MNRAS.537.3895D,2025ApJ...980..255Y}. Recently, the most probable objects that coincide with the probability regions of the neutrino events are suspected to be different types of AGN. Interestingly, these include two object types with inherently different radio properties, radio-loud blazars \citep[e.g.][]{2018ApJ...863L..10A,2018Sci...361.1378I,2018Sci...361..147I,2021ApJ...912...54R,2022MNRAS.511.4697P,2022MNRAS.509.1646S,2023MNRAS.526..347N,2023MNRAS.523.1799P,2024ApJ...965L...2J,2024ApJ...975L..30J,2024Univ...10...78K,2024PhRvD.110l3014K,2025A&A...694L..10B,2025ApJ...979....1J,2025A&A...695A.266O} with very emphasised radio jets, and Seyfert galaxies \citep{2022Sci...378..538I,2024PhRvL.132j1002N}, which are usually radio-quiet. We note that the classical radio-loudness parameter ($R$) is defined as the ratio of the $5$~GHz radio to the $4400$~\AA{} optical flux density, and sources with $R>10$ are considered radio-loud \citep[e.g.][]{1989AJ.....98.1195K}. On the other hand, the classical definitions can be misleading, and based on their physical differences, labels of jetted and non-jetted AGN are encouraged \citep[e.g.][]{2017NatAs...1E.194P}. It was, however, discussed by \citet{2024PhRvD.110l3014K} that continuous neutrino-production in AGN would not require the involvement of relativistic (radio) jets.

AGN are believed to host accreting supermassive black holes in centres of galaxies, emitting powerful radiation throughout the entire electromagnetic spectrum. A sub-class of AGN, blazars have their radio-emitting relativistic plasma jets inclined at a small angle to the line of sight of the observer \citep{1995PASP..107..803U}, have a flat radio spectrum ($ -0.5 \le \alpha \le 0; S_\nu\propto\nu^\alpha$, where $S_\nu$ is the flux density and $\nu$ the frequency), brightness temperatures exceeding the equipartition limit \citep[$T_\mathrm{b,eq} \approx 5\times10^{10}$~K,][]{1994ApJ...426...51R}, and are often associated with X-ray \citep[e.g.][]{2013MNRAS.433.2182S,2015MNRAS.452.3457G,2021ApJ...915...98P} and $\gamma$-ray emission \citep[e.g.][]{2009MNRAS.399.2041G,2015ApJ...813L..41A,2022MNRAS.516.5702O,2024MNRAS.52710168P}. Blazars also show flux density variations in the radio \cite[e.g.][]{2020AdSpR..65..756J,2023MNRAS.525.5105H}, infrared \citep[e.g.][]{2012ApJ...759L..31J,2022ApJ...927..107S,2023Ap&SS.368...18P}, optical \citep[e.g.][]{2016A&A...591A..21M,2023MNRAS.526.1687T}, X-ray \citep[e.g.][]{2018A&A...619A..93B,2023MNRAS.526.4040M}, and $\gamma$-ray \citep{2001ApJ...548..200X,2018Ap&SS.363...45L} wavebands, occasionally associated with flaring events \citep[e.g.][]{2017ApJ...846...34A,2019ApJ...880...32L,2022MNRAS.516.5702O,2023MNRAS.521.3810N}. The pc-scale structure of blazars can be studied only with observations using very long baseline interferometry (VLBI), with angular resolutions as fine as milliarcseconds (mas) at GHz frequencies. On VLBI scales, blazar jets usually exhibit compact `core' or one-sided core--jet structures \citep[e.g.][]{2019ApJ...874...43L}, as a result of Doppler enhancement of jet emission seen under a small inclination angle.

The quasar NVSS~J171822+423948 (J1718+4239) was found to exhibit flaring behaviour in the $\gamma$-ray and optical wavebands, with accompanying mid-infrared variability in its light curve, and was proposed as possibly associated with the IceCube neutrino event IC-201221A \citep{2024ApJ...965L...2J}.

With its spectroscopic redshift, $z=2.68$ \citep{2024ApJ...965L...2J},
the blazar candidate J1718+4239 is one of the highest-redshift $\gamma$-ray emitting AGN, and is the most distant neutrino-source candidate known to date. Based on its variable nature and identified flaring events in $\gamma$-ray, optical, and mid-infrared wavebands, the source fulfils some requirements as a blazar candidate, which is further supported by total-flux-density variability in its radio emission \citep{2024RNAAS...8...44P}.
To address the possible blazar nature of J1718+4239, we analysed its radio emission, utilising data from archival kpc-scale flux density measurements, as well as VLBI observations conducted for the first time. VLBI can provide the ultimate evidence for high-brightness-temperature jet emission characteristic of blazars, owing to its high-resolution imaging capability.

Throughout the paper, we considered a standard $\Lambda$ Cold Dark Matter cosmological model, with $H_0=67.3$~km\,s$^{-1}$\,Mpc$^{-1}$,  $\Omega_\mathrm{M}=0.315$, and $\Omega_\Lambda=0.685$ \citep{2014A&A...571A..16P}. At the spectroscopic redshift of J1718+4239 ($z=2.68$), $1$~mas angular distance corresponds to $8.146$~pc projected linear size \citep{2006PASP..118.1711W}.

\begin{table*}[htpb]\setlength{\tabcolsep}{4pt}\centering
    \caption{Parameters of the images and Gaussian model components fitted to the dual-frequency VLBA data of J1718+4239.}
    \label{tab:vlba}
    \begin{tabular}{ccccccccccccc}
    \hline\hline
$\nu$& \multirow{2}{*}{C} & $S$ & $R$ & $\Phi$ & $\vartheta$   & $I_\mathrm{max}$& rms & $b_\mathrm{min}$ & $b_\mathrm{max}$& P.A.& $T_\mathrm{B}$ & \multirow{2}{*}{$\delta$} \\ 
 \small\small\small\small\small\small\small\small(GHz)&  & \small\small\small\small\small\small\small(mJy) & \small\small\small\small\small\small(mas) &\small\small\small\small\small(deg) & (mas) &\small (mJy~beam$^{-1}$)  & \small($\mu$Jy~beam$^{-1}$) &\small\small\small\small (mas) & \small\small\small(mas) &\small\small ($\degr$) & \small($10^{11}$~K) & \\ \hline
  \normalsize

4.5 & c &  $84.2\pm4.5$ &       0       &        0      &$0.18\pm0.01$    & 83.5 & 27 & 1.4 & 4.1 & 1.4 & 5.7  & 11.4 \\ 
         & j &  $~~1.1\pm0.4$ & $1.48\pm0.03$ & $53.3\pm0.8$  & $0.71\pm0.05$    &       &    \\ 
6.7 & c &  $92.6\pm4.9$ &       0       &        0      & $0.13\pm0.01$   & 91.7 & 31 & 0.9 & 2.7 &2.0 &5.9  & 11.8 \\ 
         & j &  $~~1.0\pm0.2$ & $1.40\pm0.14$ & $52.4\pm4.6$  & $1.27\pm0.28$    &       &   \\
     \hline\hline\vspace{-5pt}
    \end{tabular}
    {\\{ Notes:} Col.~1 -- central frequency, Col.~2 -- model component designation: c = core, j = jet, Col.~3 -- flux density, Col.~4 -- radial distance from the core component, Col.~5 -- component position angle with respect to the core, measured from north through east, Col.~6 -- circular Gaussian model component size (FWHM), Col.~7 -- peak intensity, Col.~8 -- image noise, Col.~9--11 -- minor and major axis (FWHM) and major axis position angle of the Gaussian restoring beam, Col.~12 -- brightness temperature, Col.~13 -- Doppler factor calculated by assuming $T_\mathrm{b,int}= 5 \times 10^{10}$~K \citep{1994ApJ...426...51R}.}
\end{table*}

\section{Observations and data reduction}\label{sec:observations}

To uncover the pc-scale structure and physical properties of J1718+4239, we conducted phase-referencing VLBI observations with the U.S. Very Long Baseline Array (VLBA; project code: BP268, PI: K. Perger). The $2$-h observational run was carried out on 2024 Aug 4, in dual-band mode between $4$ and $7$~GHz. Only $8$ out of the $10$ VLBA stations \citep{1994IEEEP..82..658N} participated, with Los Alamos 
and Kitt Peak unable to observe due to hardware issues. The data were recorded at a rate of $4096$~Mbit\,s$^{-1}$ with a $2$~s integration time in left and right circular polarizations, and were correlated at the VLBA DiFX correlator \citep{2011PASP..123..275D}. Four individual intermediate frequency (IF) channels were used in dual-band setup, centred around $4.5$ and $6.7$~GHz in pairs. A total bandwidth of $128$~MHz per polarisation was used with $128$ spectral channels in each IF. The on-source times were $69$~min, $25$~min, and $9$~min for the target (J1718+4239), the phase-reference calibrator (J1709+4318, at $1\fdg72$ angular separation from the target), and the fringe-finder (J1717+4530) sources, respectively.

The amplitudes and phases of the visibility data were calibrated with  the National Radio Astronomy Observatory (NRAO) Astronomical Image Processing System \citep[\textsc{aips}, e.g.][]{1995ASPC...82..227D,2003ASSL..285..109G} package following the standard VLBA calibration guidelines.
Two individual sets of data were prepared during the fringe-fitting step. Set $A$ was acquired by interpolating fringe-fit solutions obtained for the phase-reference source (J1709+4318) to the target (J1718+4239) data, while set $B$ was produced by performing the fringe fitting directly to the target source data.  
Both calibrated data sets were exported into \textsc{uvfits} format for further analysis in the \textsc{Difmap} package \citep{1997ASPC..125...77S}, with two sub-sets in each of them for IFs~$1-2$ (central frequency $4.5$~GHz) and $3-4$ (central frequency $6.7$~GHz). We performed imaging and  phase and amplitude self-calibration cycles through hybrid mapping utilising the \textsc{clean} deconvolution algorithm \citep{1974A&AS...15..417H}. To quantitatively characterise the brightness distribution of the quasar, circular Gaussian model components were fitted to the self-calibrated visibility data \citep{1995ASPC...82..267P}. We calculated the uncertainties for the fitted model parameters following \citet{2008AJ....136..159L}, considering an additional $5\%$ flux density calibration error. 
As the target source was found bright enough for direct fringe fitting, the physical parameters and other derived properties for J1718+4239 presented in this paper were acquired from the self-fringe-fitted data set (set $B$).

To characterise the radio spectrum and light curve of the quasar, we collected available flux density measurements from the literature. Detailed information on these data is given in Appendix~\ref{appendix:radiofd}.

\section{Results}\label{sec:results}
\begin{figure}
    \centering
    \includegraphics[width=\linewidth]{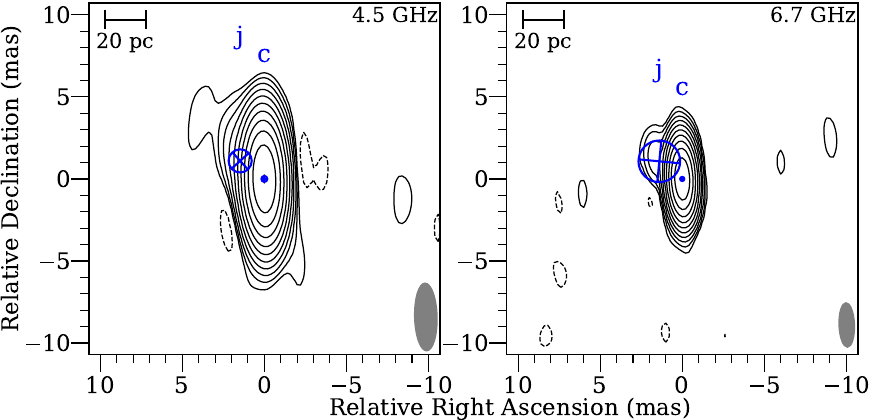}
    \caption{Naturally-weighted \textsc{clean} maps of J1718+4239 at $4.5$~GHz (IFs $1-2$) and $6.7$~GHz (IFs $3-4$). The contours start at $\pm3\sigma$ rms noise levels, and the positive contours increase by a factor of $2$. The fitted Gaussian model components are shown with blue labels and crossed circles. The angular scaling at the redshift of the quasar ($z=2.68$) is denoted at the top left corner, while the restoring beam is shown in the bottom right corner. Image parameters are listed in Table~\ref{tab:vlba}.}\label{fig:vlba}
\end{figure}

The radio maps resulting from the imaging, and the parameters of the fitted models are shown in Fig.~\ref{fig:vlba} and listed in Table~\ref{tab:vlba}.
The source was detected as a bright compact feature with an extension towards north-east at both frequencies. Two  model components were fitted to both data sets, a compact core and a secondary jet component, marked as `c' and `j', respectively, in Fig.~\ref{fig:vlba} and Table.~\ref{tab:vlba}. The separation of the two components at both frequencies agree within the errors, thus making the identification of component `j' reliable. We note that all fitted model component sizes exceed the minimum resolvable angular size \citep{2005AJ....130.2473K} of the interferometer. We found spectral indices between $4.5$ and $6.7$~GHz of $\alpha_\mathrm{c}=0.25\pm0.08$ and $\alpha_\mathrm{j}=-0.3\pm0.4$ for the core and jet components, respectively. The spectral index for the total integrated flux densities of the VLBI components is $\alpha_\mathrm{VLBI}=0.24\pm0.19$. The flat spectrum of the source supports its classification as a blazar.

The accurate coordinates of J1718+4239 were determined 
at $4.5$~GHz as $\alpha=17^\mathrm{h}18^\mathrm{m}22\fs76569$ ($\pm0.2$~mas) and $\delta=42\degr39\arcmin45\farcs0565$ ($\pm0.9$~mas). The position at $6.7$~GHz agrees with the $4.5$-GHz coordinates within the uncertainties.

\section{Discussion}\label{sec:discussion}

\subsection{Flux density variability}

The variable nature of the radio emission of the quasar at $1.4$ and $3$~GHz was briefly noted in \citet{2024RNAAS...8...44P}. With the newly available third-epoch Very Large Array Sky Survey \citep[VLASS,][]{2020RNAAS...4..175G,2021ApJS..255...30G} observational data, the brightening following the neutrino event is even more pronounced, as illustrated in Fig.~\ref{fig:lc}. There is a clear brightening at $3$~GHz, with the flux density increasing by factors of $2$ and $3$ between the first epoch (2019) and later observations (2021 and 2024), respectively. The increase in the $3$-GHz flux density might be connected to the neutrino event and the optical--$\gamma$-ray flaring detected from the AGN. Considering the angular resolution of the NRAO VLA Sky Survey \citep[NVSS; $45^{\prime\prime}$,][]{1998AJ....115.1693C}, the Faint Images of the Radio Sky at Twenty-Centimeters survey \citep[FIRST; $5^{\prime\prime}$,][]{1997ApJ...475..479W}, and the Rapid Australian Square Kilometre Array Pathfinder (ASKAP) Continuum Survey \citep[RACS; $10\farcs1$,][]{2024PASA...41....3D}, with the finer-resolution surveys showing higher flux density values, variability at $1.4$~GHz can also be recognised. As the RACS observations were conducted just $8$~days after the neutrino event IC-201221A on 2020 Dec 21, assuming that the NVSS flux density was measured when the source was in quiescence, it is possible that the brightening was temporally coincident with the neutrino event.

\subsection{Compactness}

The structure of the quasar J1718+4239 is known to be compact on arcsec scales, as seen in the large radio surveys. At $1.4$~GHz, the ratios of the peak intensity to the total flux density are $0.97$~beam$^{-1}$ and $0.94$~beam$^{-1}$ in the NVSS and FIRST surveys, respectively. Similarly, the $3$-GHz VLASS data give compactness values $0.90$~beam$^{-1}$ and $0.96$~beam$^{-1}$, for the first and third epochs, respectively. The compactness is somewhat lower at the second epoch ($0.69$).
According to our observations, the quasar is also very compact at mas scales, with the ratio of the peak intensity to the total integrated flux density being $C_\mathrm{VLBI}=0.98$~beam$^{-1}$.

As the third-epoch VLASS measurements and our VLBA observations were carried out just $2$ days apart (see Appendix~\ref{appendix:radiofd}), the kpc-to-pc compactness of the quasar can be inferred without the need to take the variability into account. By extrapolating our 4.5- and 6.7-GHz measurements, assuming the spectral index determined for the total integrated flux density of the pc-scale structure $\alpha_\mathrm{VLBI}=0.24\pm0.19$, we estimated the VLBI flux density at $3$-GHz as $S_\mathrm{3GHz,VLBI}=77.3\pm5.4$~mJy. Using this value, the flux density loss due to the resolution differences between the VLA and the VLBA could be estimated, giving a loss of $\sim15\%$ only.

\subsection{Jet parameters}

The brightness temperature of the core component was calculated following the equation \citep[e.g.][]{1982ApJ...252..102C}:
\begin{equation}
  T_\mathrm{b} = 1.22\times10^{12}(1+z)\frac{S}{\vartheta^2\nu^2}~\mathrm{K},
\end{equation}
where $S$ is the flux density measured in Jy, $\vartheta$ the full width at half-maximum (FWHM) size of the component in mas, and $\nu$ the observing frequency in GHz.
We found values of $T_\mathrm{b,4.5GHz}=(5.7\pm0.3)\times10^{11}$~K and
$T_\mathrm{b,6.7GHz}=(5.9\pm0.4)\times10^{11}$~K for the $4.5$- and $6.7$-GHz data sets, respectively (Table~\ref{tab:vlba}).  

Assuming the equipartition brightness temperature $T_\mathrm{b,int} \approx 5\times10^{10}$~K \citep{1994ApJ...426...51R} as the intrinsic value, we calculated the Doppler factors according to the relation of
\begin{equation}
  \delta=\frac{T_\mathrm{b}}{T_\mathrm{b,int}},
\end{equation}
and found values of $\delta_\mathrm{4.5GHz}=11.4$ and  $\delta_\mathrm{6.7GHz}=11.8$ from the models based on the $4.5$- and $6.7$-GHz data sets, respectively.

Following the relations \citep{1995PASP..107..803U}
\begin{equation}
\sin\theta\le\delta^{-1}\hspace{15pt}\mathrm{and}\hspace{15pt}\gamma\ge\frac{1}{2}\left(\delta+\frac{1}{\delta}\right),
\end{equation}
we estimated the limiting values for the inclination angle, $\theta$, and the bulk Lorentz factor, $\gamma$, resulting in constraints of $\theta\le5\degr$ and $\gamma\ge6$.

Our VLBI observations clearly indicate that J1718+4239 is a blazar with a relativistic jet pointing close to the line of sight, but are certainly not suitable for deciding whether this blazar is the source of the IC-201221A neutrino event or not. However, we note that a recent population study of blazars associated with IceCube neutrino events found that their VLBI-determined Doppler factors (median value around 30) are higher and jet inclination angles (median value around $0\fdg7$) are lower compared to other typical bright radio sources  \citep{2025arXiv250308667P}. Based on the limits we derived for J1718+4239, this source would fit in this population. Multi-epoch VLBI monitoring observations in the coming years could help measuring the apparent jet proper motion and refining the jet parameters in J1718+4239.

\section{Summary and conclusions}\label{summary}

To reveal the nature of the radio AGN J1718+4239, a candidate object for association with the IceCube neutrino event IC-201221A, we performed high-resolution VLBI observations in dual-band mode ($4.5$ and $6.7$~GHz) with the VLBA for the first time. We also collected total flux density measurement data from various radio sky surveys at multiple frequencies from $0.144$ to $8.4$~GHz. The source is compact ($C\approx1$) on both arcsec and mas angular scales, and shows significant flux density variability. On mas scale, we found that J1718+4239 has a compact core and a weak secondary jet component, potentially allowing for apparent jet proper motion studies in the future. This blazar is sufficiently bright and compact for direct fringe fitting, making it suitable as a phase-reference calibrator for VLBI observations of nearby weak radio sources. We derived its relative astrometric position with sub-mas accuracy. The core component has a flat spectrum, with a brightness temperature exceeding the equipartition limit ($T_\mathrm{b}\approx6\times10^{11}$~K), and a Doppler factor of $\delta\approx12$. The limiting values on the bulk Lorentz factor and the position angle were found to be $\gamma\ge6$ and $\theta\le5\degr$, respectively. These confirm that J1718+4239 is a blazar-type AGN. 

\section*{Data availability}
Raw VLBA data can be obtained from the NRAO archive\footnote{\url{https://data.nrao.edu/portal/}} under the project code BP268. The calibrated visibility data may be shared on reasonable request to the corresponding author.

\begin{acknowledgements}
We thank the anonymous reviewer for their constructive suggestions to improve our manuscript. The National Radio Astronomy Observatory is a facility of the National Science Foundation operated under cooperative agreement by Associated Universities, Inc. CIRADA is funded by a grant from the Canada Foundation for Innovation 2017 Innovation Fund (Project 35999), as well as by the Provinces of Ontario, British Columbia, Alberta, Manitoba and Quebec. 
This work made use of the DiFX software correlator developed at Swinburne University of Technology as part of the Australian Major National Research Facilities program.
We thank the Hungarian National Research, Development and Innovation Office (NKFIH, grant OTKA K134213) for support. This research was also supported by HUN-REN and the NKFIH excellence grant TKP2021-NKTA-64. E.K. acknowledges support from the German Science Foundation DFG, via the Collaborative Research Center SFB1491: Cosmic Interacting Matters -- from Source to Signal (grant no.~445052434).
\end{acknowledgements}

\bibliographystyle{aa}
\bibliography{aa55177-25.bib}

\begin{appendix}
\onecolumn
\section{Radio flux densities}\label{appendix:radiofd}
\begin{figure}[h!]
    \centering
    \includegraphics[width=0.525\linewidth]{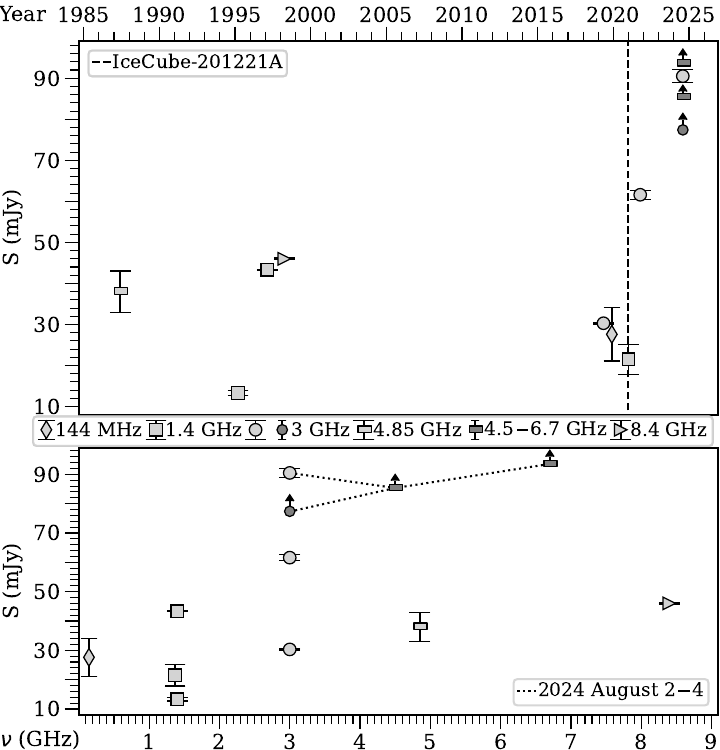}
    \caption{Radio light curve and radio spectrum of J1718+4239. The lower limits denote the VLBA flux densities. The archival observations at $0.144$, $1.4$, $3$, $4.85$, and $8.4$~GHz frequencies are indicated with light grey diamonds, squares, circles, rectangles, and triangles, respectively. Our VLBA measurements at $4.5$ and $6.7$~GHz, and the estimated $3$~GHz VLBI flux densities are denoted with dark grey rectangle and circle markers, respectively. The dashed line in the top panel marks the date of the neutrino event, while the dotted lines in the bottom panel highlight the flux densities from the coinciding VLA and VLBA observations. Detailed information is given in Table~\ref{tab:radiofd}.}\label{fig:lc}
\end{figure}

We collected data from the Low-Frequency Array (LOFAR) Two-metre Sky Survey second data release \citep[LoTSS DR2,][]{2022AnA...659A...1S} at $144$~MHz, the NRAO VLA Sky Survey \citep[NVSS,][]{1998AJ....115.1693C}, the Faint Images of the Radio Sky at Twenty-Centimeters 
\citep[FIRST,][]{1997ApJ...475..479W} survey, and the Rapid Australian Square Kilometre Array Pathfinder (ASKAP) Continuum Survey \citep[RACS,][]{2024PASA...41....3D} at $1.4$~GHz, the
 Very Large Array Sky Survey \citep[VLASS,][]{2020RNAAS...4..175G,2021ApJS..255...30G} at $3$~GHz, the
Green Bank $4.85$-GHz survey \citep[GB6,][]{1996ApJS..103..427G}, and the $8.4$-GHz Cosmic Lens All-Sky Survey \citep[CLASS,][]{2007MNRAS.376..371J,2003MNRAS.341....1M}. We note that VLASS flux densities were collected from the third and second version `quick-look' catalogues for the first and second epochs\footnote{Component catalogues are available at \url{https://cirada.ca/vlasscatalogueql0}}, respectively. As, to date, no official catalogue was released for the third-epoch observations, the flux density value was obtained by fitting a Gaussian model component to the \textsc{clean} image in \textsc{aips} using the \textsc{imfit} task. We note that, by chance, our VLBA observations (Sect.~\ref{sec:observations}) were quasi-simultaneous with the third-epoch VLASS observations. 
Detailed information on the radio observations are listed in Table\ref{tab:radiofd} and illustrated in Fig.~\ref{fig:lc}.

\begin{table}[h!]\caption{Radio flux densities of J1718+4239 from the literature and our new VLBA observations.}\label{tab:radiofd}
    \centering
    \begin{tabular}{rccccc}
    \hline\hline										
Frequency	&	Survey	&	Date	&	Flux density	&	Restoring beam	size &	Reference	\\	\hline
$144$~MHz	&	LoTSS	&	2019-11-24	&	$27.6\pm6.5$	&	$20''\times20''$	&	\citet{2022AnA...659A...1S}	\\	
$1.4$~GHz	&	NVSS	&	1995-03-12	&	$13.3\pm0.6$	&	$45''\times45''$	&	\citet{1998AJ....115.1693C}	\\	
$1.4$~GHz	&	FIRST	&	1997-02-17	&	$43.3\pm0.1$	&	$5\farcs4\times5\farcs4$	&	\citet{1997ApJ...475..479W}	\\	
$1.4$~GHz	&	RACS	&	2020-12-29	&	$21.5\pm3.7$	&	$10\farcs3\times42\farcs4$	&	\citet{2024PASA...41....3D}	\\	
$3$~GHz	&	VLASS	&	2019-05-06	&	$30.3\pm0.2$	&	$2\farcs4\times3\farcs0$	&	\citet{2020RNAAS...4..175G,2021ApJS..255...30G}	\\	
$3$~GHz	&	VLASS	&	2021-10-08	&	$61.6\pm1.1$	&	$2\farcs4\times3\farcs0$	&	\citet{2020RNAAS...4..175G,2021ApJS..255...30G}	\\	
$3$~GHz	&	VLASS	&	2024-08-02	&	$90.5\pm1.6$	&	$2\farcs2\times2\farcs8$	&	\citet{2020RNAAS...4..175G,2021ApJS..255...30G}	\\	
$4.5$~GHz	&	VLBA	&	2024-08-04	&	$85.4\pm4.9$	&	$1.4\,\mathrm{mas} \times 4.1\,\mathrm{mas}$	&	this work	\\	
$4.85$~GHz	&	GB6	&	1986-11--1987-10	&	$38\pm5$	&	$3\farcm4\times3\farcm6~^{\dagger}$	&	\citet{1996ApJS..103..427G}	\\	
$6.7$~GHz	&	VLBA	&	2024-08-04	&	$93.6\pm5.1$	&	$0.9\,\mathrm{mas} \times 2.7\,\mathrm{mas}$	&	this work	\\	
$8.4$~GHz	&	CLASS	&	1998-04-05	&	$46\pm0.2$	&	$0\farcs4-0\farcs6~^{\dagger\dagger}$	&	\citet{2007MNRAS.376..371J,2003MNRAS.341....1M}	\\	\hline\hline
    \end{tabular}\\
 {$^\dagger$ FWHM of point source response. $^{\dagger\dagger}$ X-band resolution of VLA in A configuration, considering natural weighting and snapshot observation corrections.}
\end{table}

\end{appendix}

\end{document}